\documentclass[numberedappendix]{emulateapj}
\usepackage{natbib}
\usepackage{graphicx}
\usepackage{dcolumn}
\usepackage{bm}
\usepackage{amsmath}
\usepackage{amstext}
\usepackage{pstricks}
\usepackage{color}
\usepackage{hyperref}
\newcommand{\degree}{^\circ}

\begin{document}

\title{Sharper \textit{Fermi} LAT images: instrument response functions for an improved event selection}

\author{Stephen K. N. Portillo\altaffilmark{1}, Douglas P. Finkbeiner\altaffilmark{1,2}}

\altaffiltext{1}{Harvard-Smithsonian Center for Astrophysics, 
  60 Garden Street, MS-51, Cambridge, MA 02138 USA} 

\altaffiltext{2}{Physics Department, 
  Harvard University, 
  Cambridge, MA 02138 USA}

\begin{abstract}
The Large Area Telescope on the \textit{Fermi} Gamma-ray Space Telescope has a point spread function with large tails, consisting of events affected by tracker inefficiencies, inactive volumes, and hard scattering; these tails can make source confusion a limiting factor. The parameter CTBCORE, available in the publicly available Extended \textit{Fermi} LAT data\footnote{Available at \texttt{http://fermi.gsfc.nasa.gov/ssc/data/access/}}, estimates the quality of each event's direction reconstruction; by implementing a cut in this parameter, the tails of the point spread function can be suppressed at the cost of losing effective area. We implement cuts on CTBCORE and present updated instrument response functions derived from the \textit{Fermi} LAT data itself, along with all-sky maps generated with these cuts. Having shown the effectiveness of these cuts, especially at low energies, we encourage their use in analyses where angular resolution is more important than Poisson noise.\end{abstract}

\section{Introduction}
\label{sec:intro}

The primary instrument on the \textit{Fermi} Gamma-ray Space Telescope is the Large Area Telescope (\textit{Fermi} LAT), an imaging pair conversion gamma-ray telescope. It is designed to cover an energy range $\approx 20$ MeV -- $> 300$ GeV, and has a field of view $\approx 2.5$ sr covering the sky every three hours. The instrument consists of an array of tracker sections in front of calorimeter sections. The tracker sections have 18 layers of silicon-strip detectors, with the first 16 layers behind tungsten foil in which a gamma ray may convert to an electron-positron pair, which are detected through their ionization of the silicon-strip detectors. After traversing the tracker section, the pair then lose most of their remaining energy in the calorimeter. A detailed description of \textit{Fermi} LAT can be found in \cite{2009ApJ...697.1071A}.

\textit{Fermi} LAT's point spread function (PSF) is limited to a 68\% containment radius $\mathtt{\sim} 0.1\degree$ by the lever arm of the tracker layers and the pitch of the silicon-strip detectors. At gamma-ray energies below $\approx$ 10 GeV, the PSF worsens with lower energy due to multiple scattering of the primary electron pair within the tungsten foil. The backmost 4 tungsten layers are thicker in order to improve the effective area and field of view, especially at high energy. Events that convert in these layers, so called back-converting events, have worse PSFs due to an increased chance of multiple Coulomb scattering between the first few hits of the track and their shorter tracks \citep{2007APh....28..422A}.

The \textit{Fermi} LAT Collaboration places events into analysis classes whose balance between effective area, angular resolution, and cosmic ray contamination is designed for a variety of analyses. This set of analysis classes is nested, with the more restrictive classes having smaller effective area but narrower PSFs and fewer misidentified cosmic rays. Each analysis class is further subdivided into back-converting and front-converting events, each subset being treated with a different instrument response function (IRF). We seek to further restrict these existing analysis classes by the quality of their direction reconstruction, in order to create new analysis classes with lower effective area but narrower PSFs. This work implements this classification using a cut in the parameter CTBCORE, found in the extended event data. We go on to derive IRFs for these new classes using the original IRFs and the \textit{Fermi} LAT data.

This method of improving the PSF relies on creating restrictive subsets of events, which sacrifices effective area. Thus, these new classes are most useful for analyses that are limited by source confusion rather than gamma-ray statistics. The trade off between PSF and effective area is energy dependent: at low energies, the \textit{Fermi} LAT PSF containment radii are larger and more photons are observed, making the CTBCORE cuts more useful.

\section{Methods}
\label{sec:methods}

\begin{figure}
\begin{center}
\includegraphics[width=0.49\textwidth]{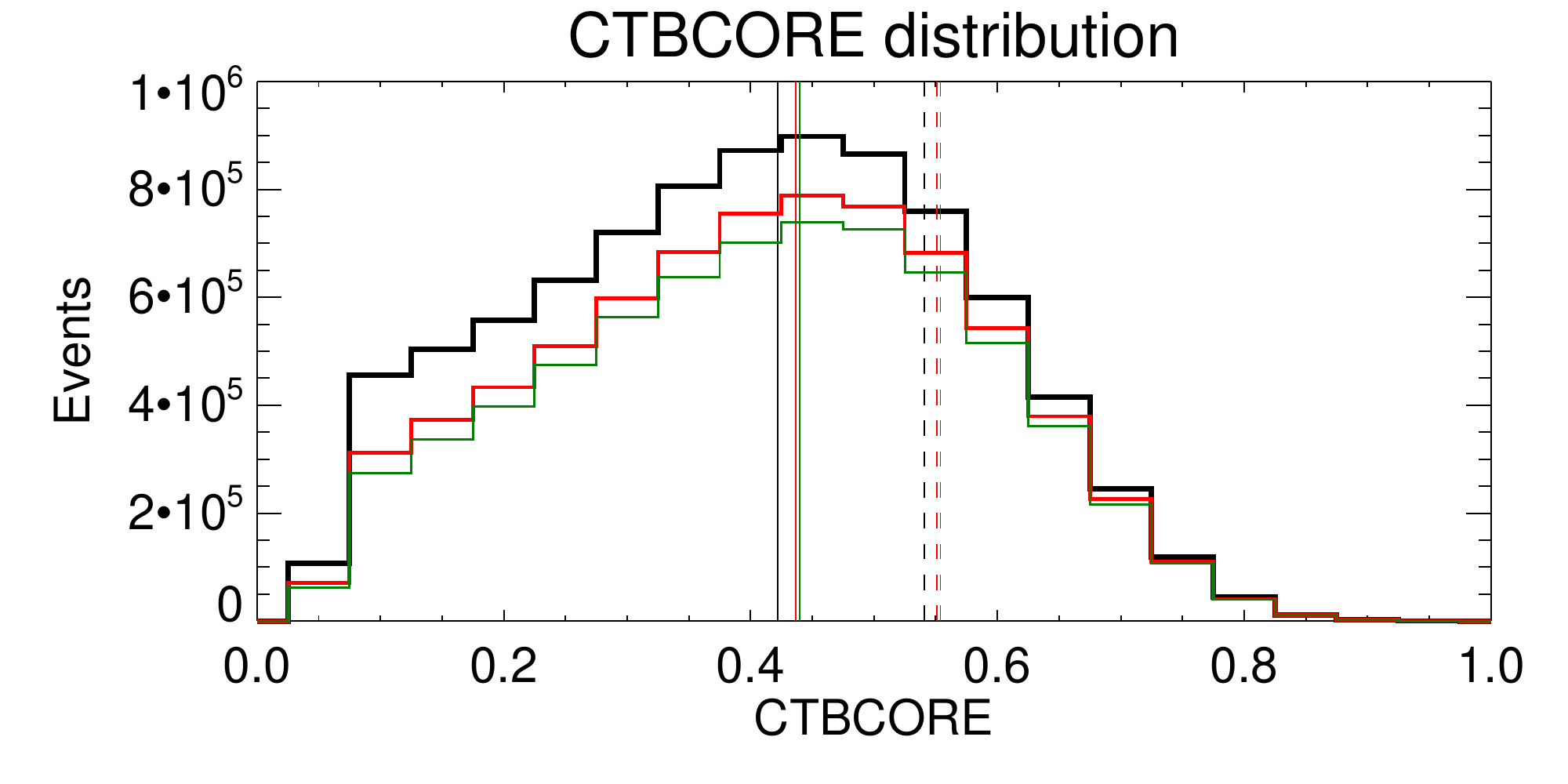}
\caption{The CTBCORE distribution for all front-converting events at energies 316 -- 562 MeV after the standard GTI and zenith angle cuts. The SOURCE, CLEAN, and ULTRACLEAN classes are the black, red, and green distributions (thickest to thinnest), respectively. The Q2 thresholds for each class are denoted by the vertical solid lines, and the Q1 thresholds are denoted by the vertical dashed lines.}
\label{fig:CTBCOREdist}
\end{center}
\end{figure}

\begin{table}
\caption{The CTBCORE thresholds used to define the Q1 cuts (top quartile) and Q2 cuts (top half) on the existing \texttt{P7REP\_V15} analysis classes.}
\label{tab:CTBCOREthres}
\begin{tabular} {| l | l | l |} \hline
\texttt{P7REP\_V15} Analysis Class & Q2 cut & Q1 cut \\ \hline
SOURCE & 0.422213 & 0.540685 \\ \hline
CLEAN & 0.436545 & 0.550919 \\ \hline
ULTRACLEAN & 0.440114 & 0.553865 \\ \hline
\end{tabular}
\end{table}

\subsection{CTBCORE}
As a part of the event reconstruction performed by the \textit{Fermi} LAT Collaboration, the accuracy of the direction reconstruction of each event is estimated. Silicon-strip inefficiencies may remove hits from an event's track, while scattering in inactive areas and hard scattering change the direction of the secondaries from the direction of the primary gamma ray. The direction reconstruction quality is estimated using a classification tree, yielding the parameter CTBCORE. CTBCORE is roughly the probability that the reconstructed direction falls within the nominal 68\% containment radius of the true direction, given by:

\[
C_{68}(E) = \sqrt{\left[c_0 \left(\frac{E}{100 MeV}\right)^{-\beta}\right]^2+c_1^2}
\]

with $c_0 = 3.3\degree$, $c_1 = 0.1\degree$, and $\beta = 0.78$ for front-converting events \citep{2012ApJS..203....4A}. Requiring a higher CTBCORE limits the tails of the PSF at the expense of effective area \citep{2009ApJ...697.1071A}. Our analysis does not require the variable CTBCORE to precisely obey the above definition, rather, we only require that higher CTBCORE is correlated with more accurate direction reconstruction.

By cutting on CTBCORE, we create new analysis classes of data with narrower PSFs but fewer photons. Since front-converting events already have much narrower PSFs than back-converting events, we consider only front-converting events. Similarly, low energy events have the largest PSF containment radii, so we focus on characterizing our cuts below 10 GeV.  The optimal thresholds are application specific and possibly energy dependent.  For simplicity, we define two CTBCORE thresholds for each existing analysis class to approximately give the top quartile of events (denoted as the Q1 cut) and top half of events (denoted as the Q2 cut) in CTBCORE \footnote{We create a Q1 and Q2 cut for each existing analysis class, eg. \texttt{P7REP\_SOURCE\_V15\_Q1}, but in the text we refer to the Q1 and Q2 cuts generically.}. In principle, the third and fourth quartiles could be used, but with lower weighting; however, to use these events, the systematic errors on the PSF tails must be well characterized. We are not confident that these errors can be sufficiently determined from the \textit{Fermi} LAT data, so we only use events that pass the CTBCORE cut. We use the same thresholds for all energies: at 300 MeV, the Q1 and Q2 cuts give 24\% and 51\% of the events respectively, with these fractions changing to 35\% and 49\% respectively at 10 GeV. To distinguish them from the Q1 and Q2 cuts, we will refer to the front-converting events of the existing analysis classes by the retronym Allfront. Figure~\ref{fig:CTBCOREdist} shows the CTBCORE distribution and Table~\ref{tab:CTBCOREthres} shows the thresholds used to define the Q1 and Q2 cuts. In order to use these new classes in science analysis, however, we must derive the associated instrument response functions (IRFs).

This work uses the Extended Pass 7 Reprocessed data with the \texttt{P7REP\_V15} IRFs and the \texttt{v9r32p5} version of the \textit{Fermi} Science Tools. The good time interval criterion used is \texttt{``DATA\_QUAL == 1 \&\& LAT\_CONFIG == 1 \&\& ABS(ROCK\_ANGLE) $<$ 52''}, and a zenith angle cut of 100$\degree$ is used to remove Earth limb photons.

\begin{figure*}
\begin{center}
\includegraphics[width=0.99\textwidth]{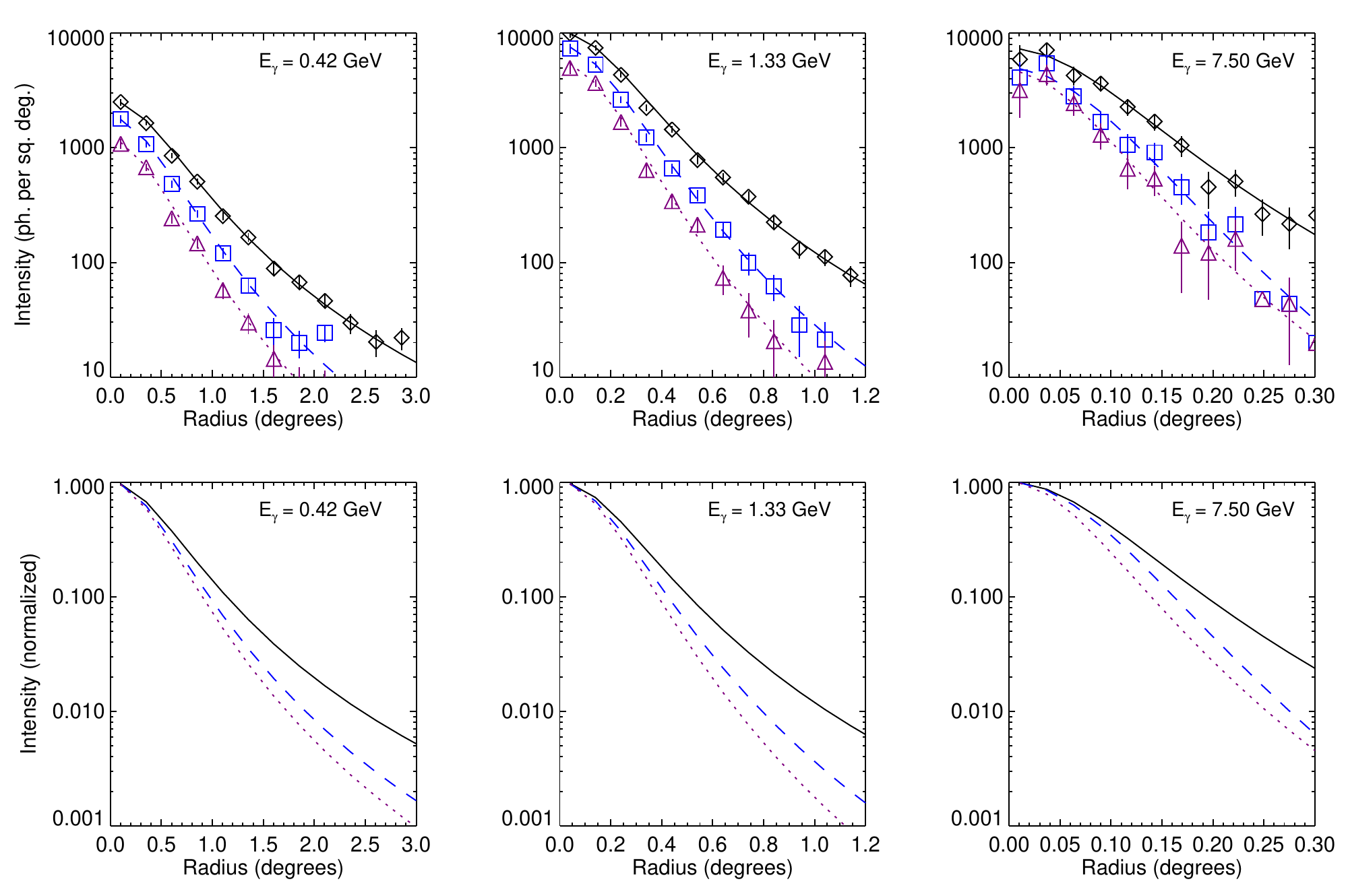}
\caption{Top row: Geminga's pulsed emission for Allfront (diamonds, black), Q2 (squares, blue), and Q1 (triangles, purple), with single King function fits (solid, dashed, and dotted lines respectively). Bottom row: single King function fits for Allfront (solid, black), Q2 (dashed, blue), and Q1 (dotted, purple), all renormalized to be unity at zero radius to emphasize the difference in the tails.}
\label{fig:psf}
\end{center}
\end{figure*}

\begin{figure}
\begin{center}
\includegraphics[width=0.49\textwidth]{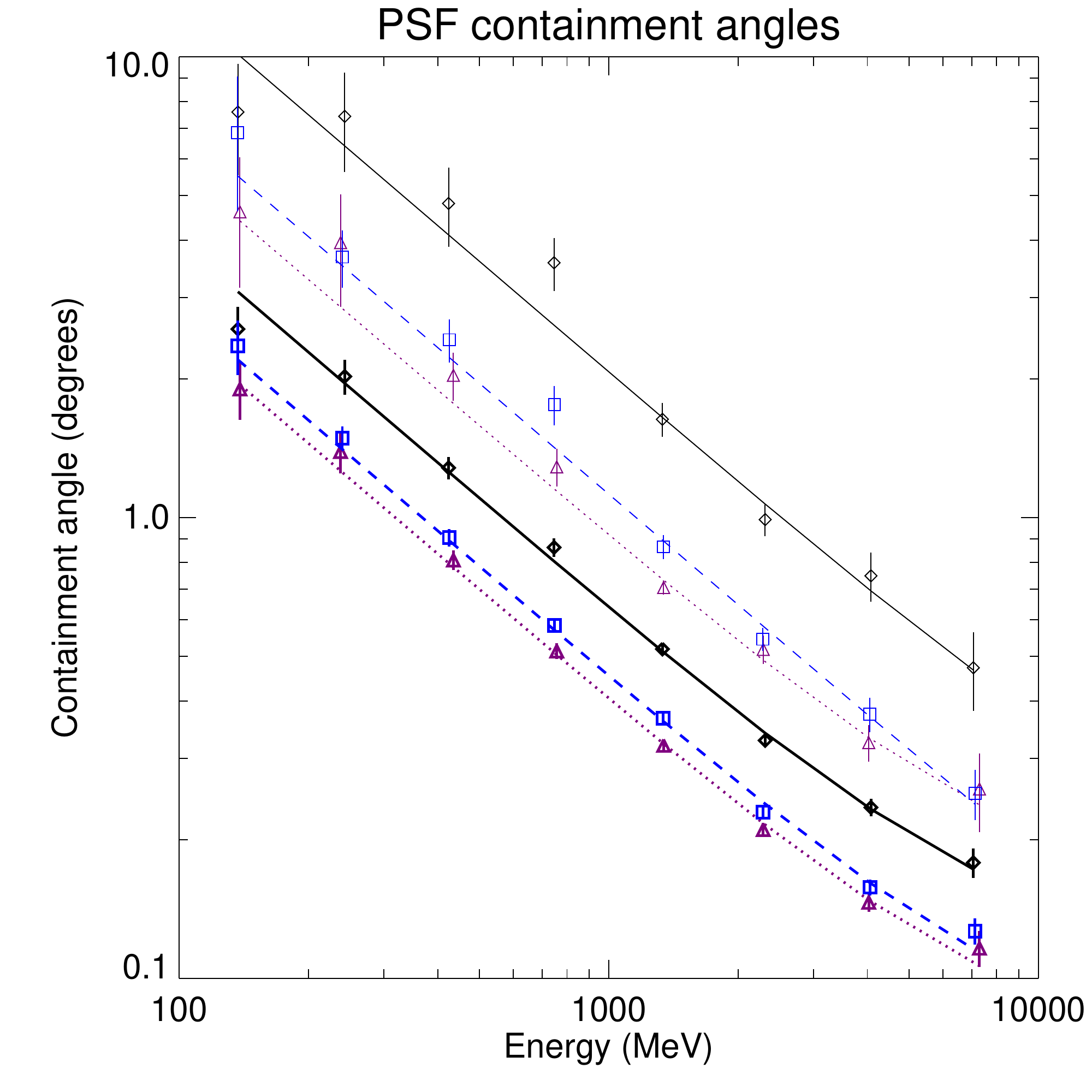}
\caption{The containment radii (68\% radius in thick lines, 95\% radius in thin lines) as determined from the Geminga pulsar for Allfront (diamonds, black), Q2 (squares, blue), and Q1 (triangles, purple), with energy-dependence smoothed fits (solid, dashed, and dotted lines respectively).}
\label{fig:psfscale}
\end{center}
\end{figure}

\begin{table}
\caption{68\% containment radii determined from the Geminga pulsar with the ULTRACLEAN analysis class, after smoothing energy dependence}
\label{tab:68radii}
\begin{tabular} {| l | l | l | l |} \hline
cut &    422 MeV&    1334 MeV&    7499 MeV \\ \hline
no CTBCORE cut&    1.4$\degree$&    0.55$\degree$&    0.19$\degree$ \\ \hline
Q2&    0.89$\degree$&    0.37$\degree$&    0.13$\degree$ \\ \hline
Q1&    0.80$\degree$&    0.33$\degree$&    0.12$\degree$ \\ \hline
\end{tabular}
\end{table}

\begin{table}
\caption{95\% containment radii determined from the Geminga pulsar with the ULTRACLEAN analysis class, after smoothing energy dependence}
\label{tab:95radii}
\begin{tabular} {| l | l | l | l |} \hline
cut &    422 MeV&    1334 MeV&    7499 MeV \\ \hline
no CTBCORE cut&    5.0$\degree$&    2.0$\degree$&    0.59$\degree$ \\ \hline
Q2&    2.2$\degree$&    0.91$\degree$&    0.33$\degree$ \\ \hline
Q1&    1.8$\degree$&    0.75$\degree$&    0.24$\degree$ \\ \hline
\end{tabular}
\end{table}

\subsection{Point Spread Function}

Following the \textit{Fermi} LAT Collaboration on-orbit PSF verification at low energy \citep{2012ApJS..203....4A, 2013ApJ...765...54A}, we determine the point spread function using the Geminga pulsar. We use the ephemeris \cite{2011ApJ...726...35A}\footnote{Available at \texttt{http://fermi.gsfc.nasa.gov/ssc/data/access/lat/ephems/}} for the pulsar phase, and accordingly restrict the data used to the range August 4, 2008 -- January 6, 2010. Half of the pulsar period is denoted off-phase (0.125--0.375 $\cup$ 0.625--0.875), and the other half on-phase. Radial profiles of the pulsar emission are created for each class in 4 energy bins per decade (100 MeV -- 10 GeV). Most of the energy dependence of the PSF is factored into the angular scale from \citep{2012ApJS..203....4A}:

\[
S_P(E) = \sqrt{\left[c_0 \left(\frac{E}{100 MeV}\right)^{-\beta}\right]^2+c_1^2}
\]

with $c_0 = 3.2\degree$, $c_1 = 0.034\degree$, and $\beta = 0.8$ for front-converting events. In each energy bin, events from within $5\times S_P(E)$ of the pulsar location are used and put into radial bins of width $0.25\times S_P(E)$. Events are not separated by incidence angle $\theta$; because of statistical limitations, we use events of all incidence angles. Thus, the PSF derived is effectively averaged over acceptance. Subtracting the off-phase image from the on-phase image removes the diffuse Galactic background and extended emission from the pulsar wind nebula, leaving an image of only the point source pulsar.

Again following the \textit{Fermi} LAT Collaboration's PSF verification, we fit this radial profile to a single King function, given by:
\[
K(x,\sigma,\gamma) = \frac{1}{2\pi\sigma^2} \left(1-\frac{1}{\gamma}\right) \left(1+\frac{x^2}{2\gamma\sigma^2}\right)^{-\gamma}
\]
Which transitions from a Gaussian core of width $\sigma$ to a power law tail $x^{-2\gamma}$. In the limit $\gamma \rightarrow \infty$, the King function exactly becomes a Gaussian of width $\sigma$. Given the large number of photons in each bin, the statistical error in each bin is approximated by Gaussian errors of variance $N_{on} + N_{off}$. Some examples of these fits are shown in Figure~\ref{fig:psf}.

Since the PSF at large radius is highly sensitive to $\gamma$, we follow the \textit{Fermi} LAT Collaboration in reparameterizing the King function in terms of its 68\% containment radius $R_{68}$ and 95\% containment radius $R_{95}$. There is no analytic mapping between $(\sigma,\gamma)$ and $(R_{68}, R_{95})$, but transforming between the two can be done numerically.

To smooth the PSF as a function of energy, the $(R_{68}, R_{95})$ in each energy bin are fit to a functional form like that of $S_P(E)$ with $c_0$ and $c_1$ refit but $\beta=0.8$ fixed (see Figure~\ref{fig:psfscale}), assuming the errors found in $(R_{68}, R_{95})$ are approximately Gaussian. Instead of taking the geometric mean energy of each energy bin, we calculate the characteristic energy of the pulsar emission in each bin. The pulsar and the diffuse emission around it have different spectra, but because the pulsar emission is brighter on-phase and the diffuse emission is not, the characteristic energy of the pulsar emission alone can be found with:

\[
\log E_{char} = \frac{\sum_{on} \log E_{\gamma} - \sum_{off} \log E_{\gamma}}{N_{on} - N_{off}}
\]

This characteristic energy deviates by less than 10\% from the geometric mean energy of the bin. Then, smoothed $(R_{68}, R_{95})$ values are calculated for the geometric mean energy in each bin using the fit parameters (some examples are given in Table~\ref{tab:68radii} and Table~\ref{tab:95radii}). From those smoothed values, $(\sigma, \gamma)$ values are determined and written to the IRF files.

\begin{figure}
\begin{center}
\includegraphics[width=0.49\textwidth]{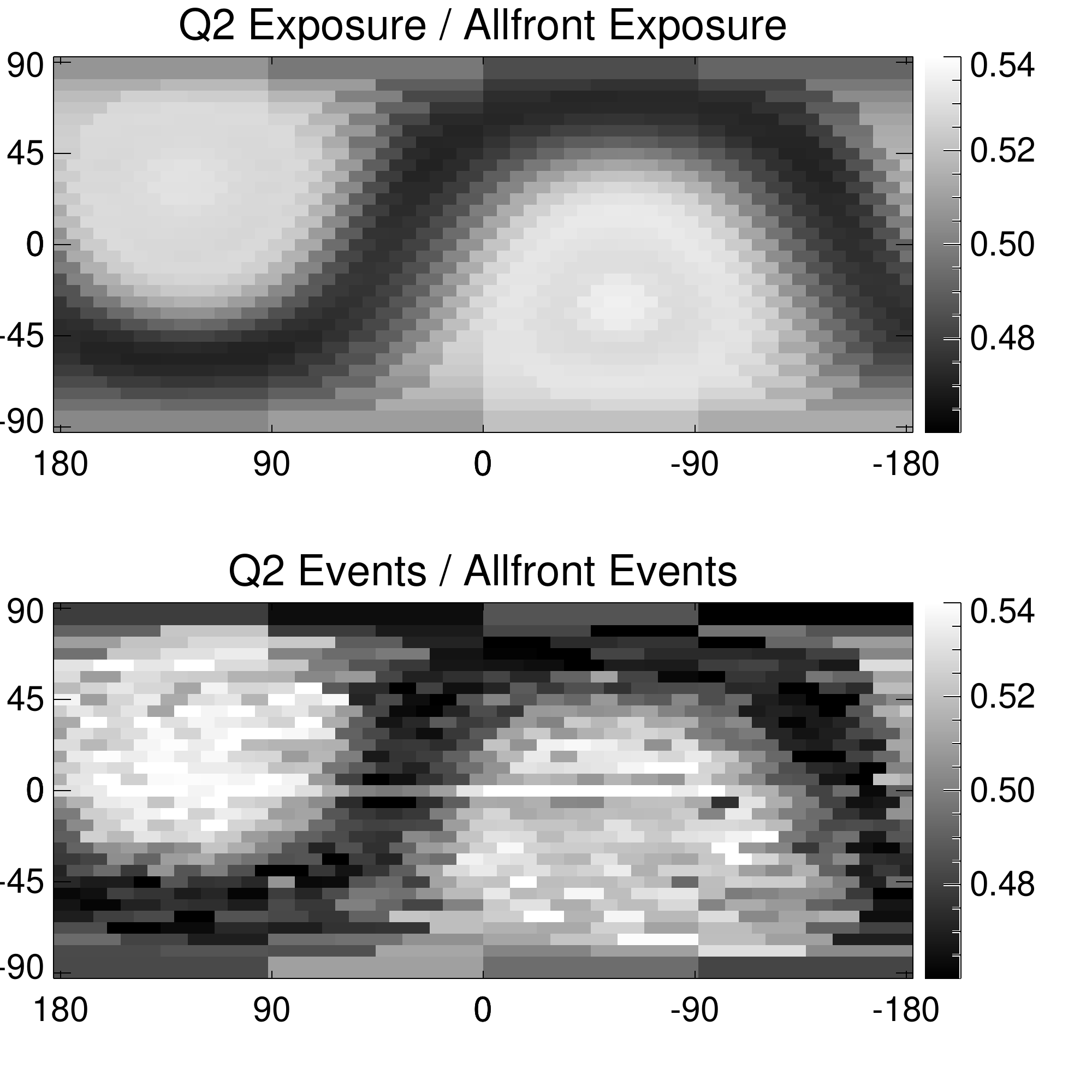}
\caption{Top: Ratio of the Q2 exposure map to the pre-cut exposure map, for the ULTRACLEAN analysis class at 365 MeV. Bottom: Surviving fraction of events after the Q2 cut, for the ULTRACLEAN analysis class at energies 316 -- 422 MeV.}
\label{fig:survfracmaps}
\end{center}
\end{figure}

\begin{figure}
\begin{center}
\includegraphics[width=0.49\textwidth]{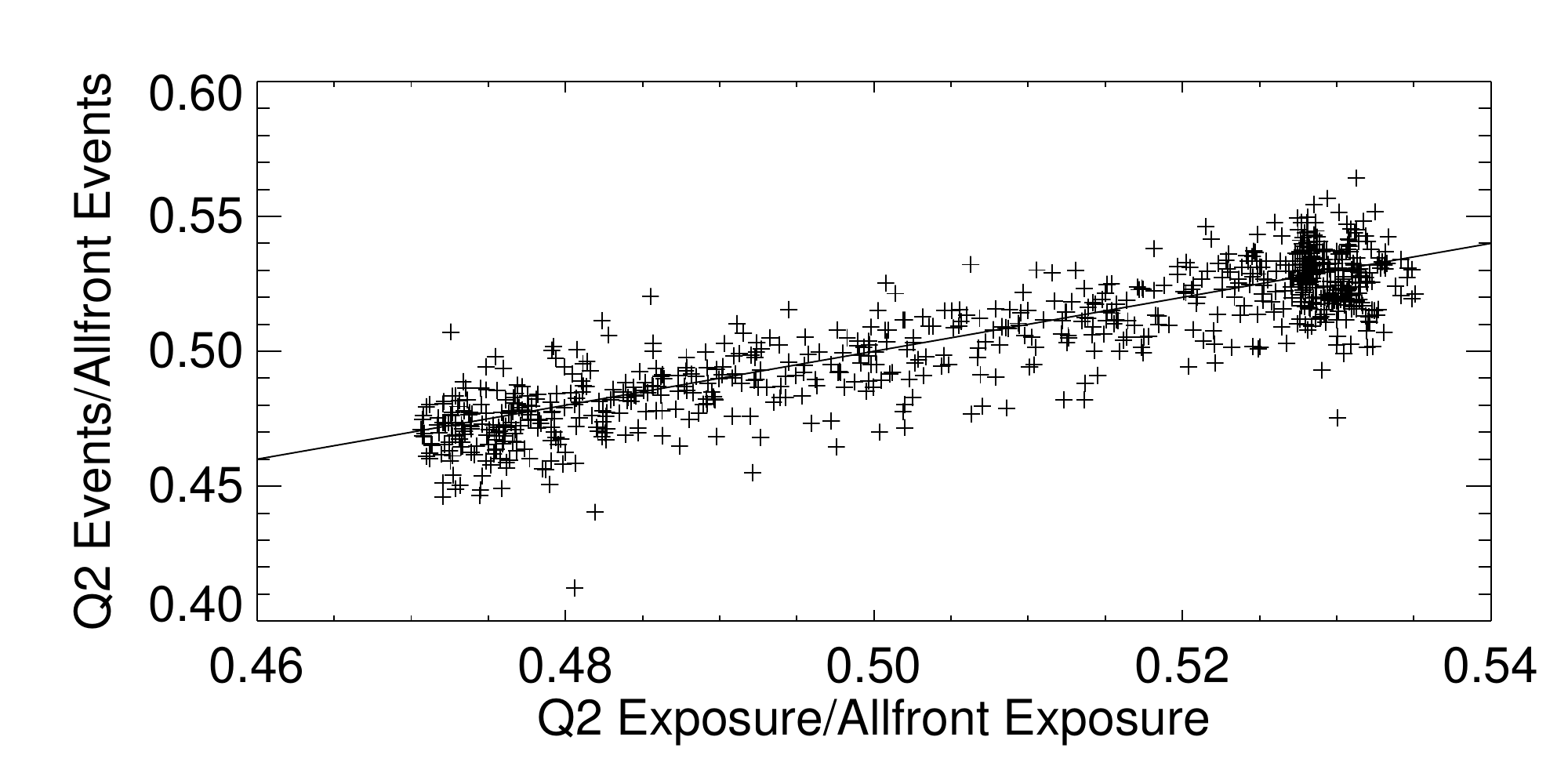}
\caption{Scatterplot of surviving fraction vs exposure map ratio for the pixels in an Nside = 8 HEALPIX map for energies 316 -- 422 MeV. The best fit slope is $a=1.0096$, with a y-offset of $b=-0.0016$ at x = 0.5.}
\label{fig:survfracscatter}
\end{center}
\end{figure}

\begin{figure*}
\begin{center}
\includegraphics[width=0.99\textwidth]{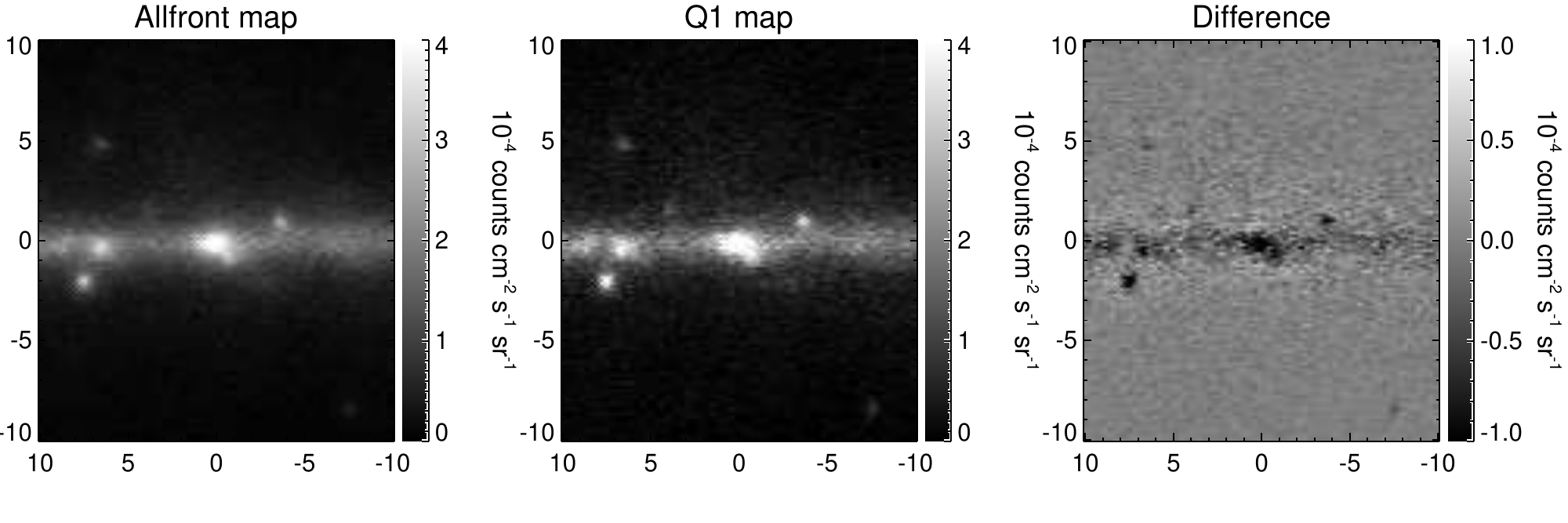}
\caption{Left: pre-cut map of the inner galaxy, Center: Q1 map of the inner galaxy, Right: pre-cut map minus Q1 map; note the deficits in the cores of bright sources and excesses in their tails.}
\label{fig:examplemaps}
\end{center}
\end{figure*}

\subsection{Effective Area}

CTBCORE's energy and incidence angle $\theta$ dependence determines the LAT's effective area for the new analysis classes. We bin all events (data from August 4, 2008 -- December 5, 2013) in an existing analysis class by energy and incidence angle, with 16 energy bins per decade and $\cos \theta$ bins of width 0.025, matching the grid that the effective area is specified on in the IRF tables provided by the \textit{Fermi} LAT Collaboration. Most bins have at least thousands of events; bins with $\cos \theta < 0.4$ are the only exception, but these incidence angles do not contribute much to the effective area. Then, the fraction of events in each bin that survive the Q1 and Q2 cuts is determined. The effective area with the new Q1 and Q2 classes at each $(E,\theta)$ bin is then taken to be the surviving fraction in that bin multiplied by the \texttt{P7REP\_V15} effective area with the existing class at that bin.

To validate the assumption that the new effective areas can be simply described by this multiplication, we use the new effective areas to predict the surviving fraction of events after the CTBCORE cuts as a function of sky position, which can then be compared with the data. In the case of uniform sky brightness, the number of events in each part of the sky is proportional to the exposure in that area. Therefore, the ratio of the Q1 or Q2 exposure map to the Allfront exposure map should equal the surviving fraction of events if the effective area has been characterized well. As can be seen in Figure~\ref{fig:survfracmaps} for the case of Q2 from 316 -- 562 MeV, the ratio of exposure maps and the surviving fraction map agree at least broadly. Events near the celestial equator are less likely to survive the CTBCORE cuts because \textit{Fermi} more often sees these areas of the sky at large incidence angles, so these events' tracks traverse more material in each tracker layer. Below $\mathtt{\sim} 1$ GeV, tracks at large incidence angles have an increased chance of multiple scattering, while above $\mathtt{\sim} 1$ GeV, hard scattering processes and additional tracker hits from the event's nascent electromagnetic shower complicate the track finding for large incidence angle events \citep{2012ApJS..203....4A}.

The Galactic Plane is seen as a feature in the surviving fraction map, but does not appear in the ratio of exposure maps; however, this behaviour is expected. A map of events can be considered to be the product of the sky's true intensity and the exposure map, convolved with the PSF of the instrument. Thus, the statement that the ratio of exposure maps is equal to the surviving fraction is slightly incorrect because of the difference in PSF pre- and post-cut. The Galactic Plane far outshines its surroundings, so events within a pre-cut PSF of the Plane are most likely from the Plane. Then, events on the Plane are more likely to survive the Q2 cut because they are from the core of the pre-cut PSF, whereas events just off the plane are less likely to survive because they are from the tail of the pre-cut PSF. Put another way, the pixels on the plane are picking up events mostly from the core of the Plane's PSF, whereas the pixels just off the plane are picking up events mostly from the tail of the Plane's PSF. Since the CTBCORE cuts prefer events from the core of the PSF, we expect a higher surviving fraction on the plane, and lower surviving fraction just off the plane, which is seen.

Even with the above consideration, we show that the ratio of exposure maps and surviving fraction maps agree quantitatively. The error in the surviving fraction in each pixel can be estimated by considering a binomial distribution with sample size equal to the pre-cut number of photons and a probability equal to the surviving fraction: for a large number of photons, the error in the surviving fraction is $\sqrt{\frac{p(1-p)}{n}}$. We try a linear fit between the surviving fraction ($y$) and ratio of exposure maps $x$, adding a noise floor $\sigma$ to reflect intrinsic scatter in the relation (see Figure~\ref{fig:survfracscatter}). Evaluating the reduced chi squared,

\[
\chi^2_R = \sum \frac{\left((y - 0.5) - a (x - 0.5) - b\right)^2}{\frac{p(1-p)}{n} + \sigma^2}
\]

we find that to get a reduced chi-squared of unity, a scatter $\sigma=0.01$ must be present. This scatter is partly due to the above consideration which makes the surviving fraction map differ from the ratio of exposure maps, but there may be other contributions from systematic shortcomings of our effective area characterization. However, since this intrinsic scatter is only 1\%, our effective area calculation is approximately correct. Ideally, CTBCORE's energy and incidence angle dependence would be studied using Monte Carlo simulations of the \textit{Fermi} LAT detector.

\subsection{Energy Dispersion}

Given that there are no astrophysical sources that give spectral lines in \textit{Fermi}'s energy range, we decide not to attempt to characterize the CTBCORE cut's effect on energy dispersion. The \textit{Fermi} LAT Collaboration characterizes the energy dispersion using Monte Carlo simulations and beam tests. In Monte Carlo simluations, the \textit{Fermi} LAT Collaboration finds a non-zero, but small, correlation between the PSF and energy dispersion: there are some events in the tails of the PSF with underestimated energies \citep{2012ApJS..203....4A}. They find this correlation to be energy and incidence angle dependent, but also find that averaged over \textit{Fermi}'s orbital procession, it causes negligible bias compared to other systematics. Thus, we take the \textit{Fermi} LAT energy dispersion as unchanged by the CTBCORE cut.

\subsection{Data Release}

The instrument response functions described above can be found at \texttt{http://skymaps.info/}, and should be used with the Extended Pass 7 Reprocessed data. These files can be used with the \textit{Fermi} Science Tools in creating exposure maps and conducting likelihood analyses. However, the \textit{Fermi} Science Tools cannot currently be used to select events based on CTBCORE, thus, events satisfying the cut must be selected with other software, like NASA HEASARC's FTOOLS.

We are also releasing full-sky maps generated from \textit{Fermi} LAT data weeks 9--288. The exposure maps used to generate these sky maps were created with the \textit{Fermi} Science Tools, using the IRFs we are releasing. Our maps are optionally point-source masked with the 1FGL: the brightest 200 sources are masked, and the rest subtracted. The maps are also optionally smoothed to a Gaussian FWHM of $2\degree$. We provide two sets of energy bins: 12 logarithmically spaced bins for use in spectral analyses (``specbin''), and 8 round-number bins for visual inspection (``imbin''). The maps are constructed in the HEALPix pixelization with $Nside=256$.

\section{Possible Applications}

Analyses looking at diffuse emission near the Galactic plane could benefit from the CTBCORE cut. The Galactic plane greatly outshines any other diffuse emission, so emission from the Galactic plane can easily leak into the region of interest at low energy. Figure~\ref{fig:examplemaps} illustrates how the CTBCORE cut reduces this leakage. At worst, a systematic excess of diffuse emission may be inferred in the region of interest because of this leakage; at best, modelling the leakage introduces a systematic uncertainty from the uncertainty of the tails of the PSF. As well, analyses involving point sources in the Galactic plane will benefit because sources so closely spaced that their PSFs overlap at low energy. \cite{2014arXiv1402.6703D} uses our method because they look at both diffuse emission just above and below the plane, as well as emission in the Galactic Center where many point sources are present.

\section{Conclusions}
\label{sec:conclusions}
The CTBCORE cuts substantially suppress the tails of the \textit{Fermi} LAT PSF by removing events with poor direction reconstructions. In the lowest energy range considered (100--178 MeV), the 95\% containment radius for the \texttt{P7REP\_SOURCE\_V15} class is $9.7\degree$, and introducing the Q2 cut (the best half of photons as ranked by CTBCORE) reduces this radius to $8.4\degree$ and introducing the Q1 cut (the best quartile of photons as ranked by CTBCORE) reduces it further to $5.8\degree$. At the highest energies considered (5623--10000 MeV), the effect is still substantial: without the CTBCORE cut, the \texttt{P7REP\_SOURCE\_V15} 95\% containment radius is $0.34\degree$, which is reduced to $0.27\degree$ by the Q2 cut, and reduced further to $0.24\degree$ by the Q1 cut. Since the CTBCORE cuts sacrifice effective area in favour of a narrower PSF, they are beneficial for analyses where source confusion is more important than Poisson statistics. At low energy, the PSF is the widest and more events are observed, so the CTBCORE cuts are most valuable at these energies. The new instrument response functions under these cuts have been characterized using the \textit{Fermi} LAT data itself: the PSF was characterized using the Geminga pulsar as a point source, and the effective area was characterized by finding the surviving fraction of events under the CTBCORE cuts in the entire \textit{Fermi} LAT dataset. These cuts are useful in any work where source confusion is an important systematic error. We have released the instrument response files necessary to use these cuts, as well as full-sky maps made with these cuts. It is our hope that similar analysis classes of events, created to address the PSF directly, may become part of the \textit{Fermi} LAT Pass 8 data release.

\vskip 0.15in {\bf \noindent Acknowledgments:} 

We would like to thank Simona Murgia and Bill Atwood for suggesting the use of the CTBCORE cut. We would also like to thank Eric Charles and Seth Digel for helpful comments. Finally, we would like to thank the \textit{Fermi} LAT Collaboration for providing the
data used in this analysis. DPF is supported in part
by the NASA Fermi Guest Investigator Program. This research made use of the NASA Astrophysics Data System (ADS) and the IDL
Astronomy User's Library at Goddard.\footnote{Available at
  \texttt{http://idlastro.gsfc.nasa.gov}}

\bibliography{hires}{}

\end{document}